\documentclass[preprint,preprintnumbers]{revtex4}

\usepackage{graphicx}

\begin{document}

\title {A green's function approach for surface state photoelectrons in topological insulators} 

\author{ D. Schmeltzer}
\affiliation{Physics Department, City College of the City University of New York,  
New York, New York 10031}


\begin{abstract}

The topology of the surface  electronic states   is   detected  with  photoemission.  We explain the photoemission from the topological surface state . This is done by identifying the effective  coupling between surface electrons-photons and vacuum electrons. The  effective electron photon coupling is given  by $e\tau^2$ where $\tau$ is the dimensionless tunneling amplitude of the zero mode surface states to tunnel into the vacuum.
We compute the polarization and   intensity of the emitted photoelectrons. We introduce a model which takes in account the Dirac Hamiltonian  for    the surface electron to  photons coupling and the tunneling of the zero mode into the vacuum.    
 Within the Green's function formalism we obtain   exact  results for the emitted  Photoelectrons to   second order in the laser field.  The  number  of  the emitted    photoelectrons  is sensitive to the  laser coherent state  intensity,   the   polarization is sensitive to    the   surface topology  of the electronic states    and the  incoming   photon  polarization.  The calculation  is performed for the helical , Zeeman and   warping case allowing to study spin  textures.

\end{abstract}

\maketitle
\textbf{I. Introduction}

\vspace{0.2 in }

Photoemission, photoconductivity, optical conductivity  and scanning tunneling microscopy  are sensitive  to the nature  of  surface states.
Photoemission is  studied using a high power  laser-based light source.  It has been  shown that the spin polarization of the  photoelectrons  emitted from the surface of    Bi$_{2}$Se$_{3}$ topological insulator $(TI)$ \cite{Volkov,Zhang,Kane,David}  can be manipulated  through the  laser light  polarization \cite{Nature,xue} finds that the photoelectron  polarization  is completely  different  from the initial state and is controlledr by the  photon polarization. Few  explanation based  on  phenomenological models have been proposed \cite{Park,Wang,Moore}. 
In a recent photoemission experiment \cite{Photo} the authors have demonstrated $\pm$ 100 \% reversal of a single component of the measured spin polarization vector upon the rotation of light polarization as well as full three-dimensional manipulation by varying experimental configuration and photon energy. This experiment shows that the photoelectrons spin polarization is achievable in systems with a layer-dependent, entangled spin-orbital texture. 
There are also other studies of spin-polarized photoelectrons spectroscopy of TI \cite{suga} including orbital-selective spin textures \cite{xie} and  reports  that  interactions  \cite{Lin} might affect the photoemission spectrum.
Regarding the helicity-dependent photocurrent, due to the spin selection rules one finds that circularly polarized light  excites the surface states and an electric DC current  is observed  \cite{Steinberg} and was investigated by  \cite{Oppen}. The authors compute  the induced DC current using the TI surface model for Bi$_{2}$Se$_{3}$ which also includes  the warping nonlinearity and the presence of the Zeeman magnetic term \cite{Oppen}.

In spite of this success the main question  of how to  explains photoemmission for Topological Insulators $TI$ remains open.  It is not clear  what is the electron-photon coupling  which describes the coupling  between the electrons in the vacuum and in  the solid.
On the surface of the $TI$ the  coupling  is given by the Dirac form $\vec{\sigma}\cdot \vec{A}$ and  in the vacuum by  $\vec{A}\cdot\vec{p}$, as a result neither Hamiltonians can describes the transition between the surface and vacuum states.
Physically one describes photoemmision as a process where a photon is absorbed and an electron is excited from the surface  to the vacuum, clearly no such matrix element exists for the surface electrons (for the bulk electrons the situation is different since   the electron photon coupling is given by $\vec{A}\cdot \vec{p}$  and the matrix element $\langle vacuum-electron \vec{A}\cdot \vec{p} bulk -electron \rangle \neq 0$). 
Such a process require the knowledge of the electron photon vertex,  $surface (electron)-photon-vacuum(electron)$. 
We solve this problem  using  the zero mode  surface $TI$ state.  The surface state are localized at $ z=0$  with  the  amplitude  $e^{-\kappa z}$ \cite{Fan}  ($ z<0$ represents the solid and $ z>0$ describes the vacuum, the eigenstates on the surface  have a  small  amplitude to tunnel  into the vacuum.) $z=0$ represents the surface and   $z>d$  describes  the free  electrons      separated by the surface-vacuum  binding  energy $V_{0}$.

The effective coupling is given by $e\cdot \tau^2$ ($e$ is the charge and  $\tau$ is the overlap between  the two types of wave functions).

 We find that the polarization measured by the detector  depends on the  product :  the projection of the  electron spin polarization on  the direction of 
the detector,  the scalar product between the photon vertices (which result from the spinor form of the electron operator) and the transverse polarization of the incoming photon) and  the intensity which  measures the number of the photoelectrons emitted. 
 
The plan of this paper is  as follows.  In Sec. II we present  the  model   for  surface    in the presence of the photon field and tunneling  amplitude into the vacuum.   In Sec. III  we   introduce the Green's function and compute the number of the photoelectrons emitted for the helical, Zeeman and  warping case.

Section IV  deals with the study of  the polarization  for the helical, Zeeman and  warping case as function  of the polarization and intensity of  the incoming  photons.  Section V is devoted to discussions  and  presents  our main  conclusions.

\vspace{0.2 in} 

\textbf{II- Model for  the   photoemission from a TI surface} 

\vspace{0.2 in}
  
\noindent The photoemission   for the $TI$ involves a   four component spinor for  the bulk, a  two component spinor for the surface  and  a wave function for the vacuum.
The surface electrons  at $z=0$  have an amplitude $\frac{e^{-\kappa z}}{\sqrt{N(\kappa,\kappa_{0})}}$ to tunnel  into the vacuum ($z>0$). The  electrons  detected by the detector have  a mean free path of $L$. To simplify the problem we expand  in plane  waves   $\frac{e^{i k_{z} z}}{\sqrt{L}}$ the 
vacuum electrons using a box of length $L$. We have for the vacuum electron, $b_{\sigma}(\vec{K},z>0)=\sum_{k_{z}}\frac{e^{i k_{z} z}}{\sqrt{L}} b_{\sigma}(\vec{K},k_{z}) $.
     The electrons on the surface 
are described by the spinor   $\Psi(\vec{r}{\perp})$  with  the eigenspinor $|u(\vec{K})\rangle$,  $\Psi(\vec{r}_{\perp})=\int\frac{d^2 K}{(2\pi)^2}e^{i\vec{K}\cdot \vec{r}_{\perp}}C(\vec{K})u(\vec{K})$, $u(\vec{K})$ is the two component spinor $ U_{\alpha=\uparrow}(\vec{K})=\langle \uparrow |u(\vec{K}) \rangle$ and    $U_{\alpha=\downarrow}(\vec{K})=\langle \downarrow |u(\vec{K}) \rangle$.
For $z>0$ we have  $\Psi(\vec{r}_{\perp},z)\approx \frac{e^{-\kappa z}}{\sqrt{N(\kappa,\kappa_{0})}} \Psi(\vec{r}_{\perp},z=0)$ and  for  $z<0$  we have $\Psi(\vec{r}_{\perp},z)\approx \frac{e^{\kappa_{0} z}}{\sqrt{N(\kappa,\kappa_{0})}} \Psi(\vec{r}_{\perp},z=0)$.
 The  surface electrons overlap  the with the vacuum  electrons  $\frac{e^{i k_{z} z}}{\sqrt{L}}$ in region  closed to z=0. We restrict    the overlap to  the region  $-d_{0}\leq z\leq d$ where  $d_{0}=\frac{1}{\kappa_{0}}$ and  $d= \frac{1}{\kappa}$ and find the tunneling matrix element $t(k_{z})$.
($\frac{1}{\kappa_{0}}$ inverted gap of the $TI$ and $\frac{1}{\kappa_{0}}$ is gap of the vacuum and $d$, $d_{0}$ are   a  few lattice constants).
\begin{eqnarray}
&&\int_{-d_{0}=-\frac{1}{\kappa_{0}}}^{d=\frac{1}{\kappa}}\,dz C^{\dagger}(\vec{K},z)u_{\sigma}^{*} (\vec{K}) b_{\sigma}(\vec{K},z)+h.c.\nonumber\\&&= 
C^{\dagger}(\vec{K},z=0)u_{\sigma}^{*} (\vec{K})\int_{-d_{0}=-\frac{1}{\kappa_{0}}}^{d=\frac{1}{\kappa}}\,dz \frac{e^{-\kappa(z) z}}{\sqrt{N(\kappa,\kappa_{0})}} \frac{e^{i k_{z} z}}{\sqrt{L}} \sum_{k_{z}} b_{\sigma}(\vec{K},k_{z})+h.c.\nonumber\\&&
 = C^{\dagger}(\vec{K},z=0)  u_{\sigma}^{*} (\vec{K}) \sum_{k_{z}} t(k_{z})b_{\sigma}(\vec{K},k_{z})+h.c.
\nonumber\\&&
\end{eqnarray}
where  the tunneling matrix is  $t(k_{z})$  is given by ,
\begin{equation}
t(k_{z}) =\int_{-d_{0}=-\frac{1}{\kappa_{0}}}^{d=\frac{1}{\kappa}}\,dz \frac{e^{-\kappa(z) z}}{\sqrt{N(\kappa,\kappa_{0})}} \frac{e^{i k_{z} z}}{\sqrt{L}} 
\label{tun}
\end{equation}
 $t(k_{z})$ is  controlled by $\kappa(z)$, $\kappa(z>0)=\kappa$, $\kappa(z<0)=\kappa_{0}$ and the normalization factor $\frac{1}{\sqrt{N(\kappa,\kappa_{0})}}$

\noindent  Using Eq.$(1)$   we  introduce the following  model: 
\begin{eqnarray}
&&H=H^{(surf.)}+H^{(vac.)} +H^{(surf.-vac.)}+H^{(ext.)}(t)\nonumber\\&&
H^{(surf.)}=\int\,\frac{d^2 K}{(2\pi)^2} \Psi^{\dagger}(\vec{K},z=0) [\hbar v(\sigma^{1}K_{2}-\sigma^{2}K_{1}) +\sigma^{3}\Delta(\vec{K},k_{c})]\Psi(\vec{K},z=0)\nonumber\\&&
H^{(vac.)}=\int\,\frac{d^2 K}{(2\pi)^2}\sum_{k_{z}}\sum_{\sigma=\uparrow,\downarrow}\left(\frac{\hbar^2 (K^2+k^2_{z})}{2 m}+V_{0}-\mu\right)b_{\sigma}^{\dagger}(\vec{K},k_{z})b_{\sigma}(\vec{K},k_{z})\nonumber\\&&
H^{(surf.-vac.)}=\int\,\frac{d^2K }{(2\pi)^2} C^{\dagger}(\vec{K},z=0)  \sum_{\sigma=\uparrow,\downarrow}u_{\sigma}^{*} (\vec{K}) \sum_{k_{z}} t(k_{z})b_{\sigma}(\vec{K},k_{z})+h.c.]  \nonumber\\&&
H^{(ext.)}(t)=(-ev)\int\,\frac{d^2K}{(2\pi)^2}\int\,\frac{d^2Q}{(2\pi)^2}C^{\dagger}(\vec{K}+\vec{Q},z=0;t)C(\vec{K},z=0;t)  \vec{W}(\vec{K}+\vec{Q},\vec{K})\cdot \vec{A}(\vec{Q},q_{z};t)\nonumber\\&&
\end{eqnarray}
\noindent $H^{(surf.)}$ describes the $TI$  surface electrons.
 We will consider three different cases:
 
\textbf{A-}  The helical state  $\mathbf{\Delta(\vec{K},k_{c})}=0$ is zero,  the eigenvalues are  given by  
$\epsilon=\epsilon_{surf.}=\epsilon_{0}= v|\vec{K}|$,
 $\hat{\epsilon}=\epsilon_{0}-\mu\equiv v|\vec{K}|-\mu$
 and the    spinors   for $|\vec{K}|\neq0$ are  ,$|u(\vec{K})\rangle=\frac{1}{\sqrt{2}}|\vec{K}\rangle\otimes[1,-ie^{i\chi(\vec{K})}]^{T}$, $e^{i\chi(\vec{K})}\equiv \frac{K_{1}+iK_{2}}{|\vec{K}|}$. The electromagnetic  vertex functions  are given by the  Pauli matrix elements,  
 $W_{1}(\vec{K}+\vec{Q},\vec{K})=\langle u((\vec{K}+\vec{Q})|(-\sigma^2 )|u(\vec{K})\rangle= \cos[\frac{1}{2}(\chi(\vec{K}+\vec{Q})+\chi(\vec{K}))]$ and
 $W_{2}(\vec{K}+\vec{Q},\vec{K})=\langle u(\vec{K}+\vec{Q})|\sigma^1 |u(\vec{K})\rangle=-\sin[\frac{1}{2}(\chi(\vec{K}+\vec{Q})+\chi(\vec{K}))]$.

\textbf{B-} For  the  Zeeman gap $\Delta\neq 0$    the eigenvalue are  given by  $\hat{\epsilon}= \hat{\epsilon}_{0}+ \Delta^2$ and the spinors are,
$|u(\vec{K},\Delta)\rangle=|\vec{k}\rangle\otimes\left[\cos\left(\frac{\beta(\vec{k})}{2}\right),-   ie^{i\chi(\vec{k})} \sin\left(\frac{\beta(\vec{k})}{2}\right)\right]^{T}$, $\cos\left[\beta(\vec{K})\right] =\frac{\Delta}{\sqrt{(\hbar v K)^2+\Delta^2}}$.   The vertex functions are for this case, 
$W_{1}(\vec{K},\vec{K}) = \sin[\beta(\vec{K})] \sin[\chi(\vec{K})]$ , $W_{2}(\vec{K},\vec{K}) = -\sin[\beta(\vec{K})] \cos[\chi(\vec{K})]$.

\textbf{C-} For the  nonlinear warping  the eigenvalue are,
 $\hat{\epsilon}= \hat{\epsilon}_{0}\Big[1+ \Big(\frac{\hat{\epsilon}_{0}}{\epsilon_{c}}\Big)^{4}\cos^2[3 \chi]\Big]^{\frac{1}{2}}$ ($\epsilon_{c}= v k_{c}$ is the warping energy.)
 The spinors are  given by,
$|u^{(+)}(\vec{K},k_{c})\rangle=|\vec{K}\rangle\otimes\left[\cos\left(\frac{\tilde{\beta}(\vec{K})}{2}\right),-   ie^{i\chi(\vec{K})} \sin\left(\frac{\tilde{\beta}(\vec{K})}{2}\right)\right]^{T}$. The vertex function take the form,
$W_{1}(\vec{K},\vec{K}) = \sin[\beta(\vec{K},k_{c})] \sin[\chi(\vec{K})]$ , $W_{2}(\vec{K},\vec{K}) = -\sin[\beta(\vec{K},k_{c})] \cos[\chi(\vec{K})]$ with 
 $\cos[\beta(\vec{K},k_{c})] = \frac{( \frac{K}{k_{c}})^2 cos[3\chi(\vec{k})] }{\sqrt{ 1+(\frac{K}{k_{c}})^4 \cos^2[3\chi(\vec{k})]}}$.

The vacuum Hamiltonian $H^{(vac.)}$ 
is controlled by the  binding energy $V_{0}$ and  eigenvalues 
$\hat{E}= E-\mu=\frac{\hbar (K^2+k^2_{z})}{2m}+V_{0}-\mu$.
The vacuum electrons operators   obey the  momentum expansions    $b_{\sigma}(\vec{K},z>0)=\sum_{k_{z}}\frac{e^{i k_{z} z}}{\sqrt{L}} b_{\sigma}(\vec{K},k_{z})$ where $L$  is the distance from the $TI$ surface  to  the  detector.

 $H^{(ext.)}(t)$  is the  electron-photon  Hamiltonian restricted to the surface at $ z=0$. $A_{i}(\vec{Q},q_{z};t)$, $W_{i}(\vec{K}+\vec{Q},\vec{K})$ are   the photon field   and  vertex for the   $i=1,2$ direction.

\noindent The high intensity photon field $\vec{A}(\vec{Q},q_{z})$  is    a coherent state $|\Omega\rangle$. The   direction of the incoming  photon  with respect to the surface  at $z=0$  is given by the vector $\vec{p}\equiv  \vec{Q}+\vec{q}_{z}$.  The two transverse linear  polarizations are   given by the vectors $\vec{e}_{s=1}(\vec{p})$ and  $\vec{e}_{s=2}(\vec{p})$  which obey $\vec{e}_{s}(\vec{p})\cdot \frac{\vec{p}}{|\vec{p}|}=0$, $\vec{e}_{s=1}(\vec{p})\cdot\vec{e}_{s=2} \vec{p}=0$.
\begin{eqnarray}
&& \vec{A}(\vec{r},t)=\sqrt{\frac{\hbar}{\tilde{\epsilon}}}\int\,\frac{d^3 p }{(2\pi)^3}\frac{1}{\sqrt{2 \Omega(p)}}\sum_{s=1,2}\Big[e^{i\vec{p}\cdot \vec{r}}\vec{e}_{s}(\vec{p})a_{s}(\vec{p})e^{-i\Omega t}+
e^{-i\vec{p}\cdot \vec{r}}\vec{e_{s}}(\vec{p})a^{\dagger}_{s}(\vec{p)}e^{i\Omega t}\Big],\nonumber\\&&
\frac{ \vec{p}}{|\vec{p}|}= (\sin(\theta)\cos(\phi),\sin(\theta)\sin(\phi),\cos(\theta)),\nonumber\\&&  
\vec{e}_{s=1}(\vec{p})=(\cos(\theta)\cos(\phi),\cos(\theta)\sin(\phi),-\sin(\theta)),\hspace{0.05 in}
\vec{e}_{s=2}(\vec{p})=(-\sin(\phi),\cos(\phi),0)\nonumber\\&&  
\end{eqnarray}
$\theta$, $\phi$ are the  photon polarization angles, $|\Omega \rangle$ is the   the lase coherent state   and $\tilde{\epsilon}$ is the dielectric constant. 

\vspace{0.2 in}

\textbf{III- The spin detection}

\vspace{0.2 in}

The detector is in the  plan $x-y$ parallel to the $TI$ surface   and perpendicular to the the $z$ axis . The detector measures the  spin polarization in the $y$ direction. When the $TI$  surface  rotates around the $y$ axes  the detector measures the angle  $\phi_{d}$ which coincide  with the spinor angle $\chi[\vec{K}] \equiv \phi_{d}$. The rotation of the sample around the $x$ axes allows to measure the momentum in the $z$ direction. We have $|\vec{K}|=|\vec{k}| \cos[\theta_{d}]$ and $k_{z}= |\vec{k}|\sin[\theta_{d}]$ where $\vec{k}=\vec{K}+\hat{z}k_{z}$. The eigenvalue of the vacuum electrons can be written in term of $\epsilon_{0}$ (the surface eigenvalue ) and $\theta_{d}$, $\hat{E}= E-\mu=\frac{\epsilon^{2}_{0}}{2 m v^{2} \sin^2[\theta_{d}]}+V_{0}-\mu$.

The spin density  $\langle \mathbf{n}^{y}\rangle$  measured by the detector is given in terms of  the Green's function $G_{\beta,\alpha}(\vec{Q},q_{z};t,t+\delta t)$:
\begin{eqnarray}
&&\langle \mathbf{n}^{y}(z=L )\rangle=-i\sum_{\alpha,\beta=1}^{\alpha,\beta=2}\sigma^{y}_{\beta,\alpha}\int_{-\infty}^{\infty}\frac{d^2 K}{(2\pi)^2}\int\frac{dk_{z}}{2\pi}
G_{\beta,\alpha}(\vec{K},k_{z};t,t+\delta t)\nonumber\\&&=(-i)\sum_{\alpha,\beta=1}^{\alpha,\beta=2} \sigma^{y}_{\beta,\alpha}\int\,\frac{d^2 K}{(2\pi)^2}\int\,\frac{dk_{z}}{2\pi}\int\frac{d\omega}{2\pi}e^{i\omega \delta t}G_{\beta,\alpha}(\vec{K},k_{z};\omega),\delta t \rightarrow 0 \nonumber\\&&
G_{\beta,\alpha}(\vec{K},q_{z};t,t+\delta t)= -i\langle g|T(b_{\beta}(\vec{K},k_{z};t)b^{\dagger}_{\alpha}(\vec{K},k_{z};t+\delta t))| g \rangle \nonumber\\&&
=-i\langle O\otimes \Omega|T\Big(b_{\beta}(\vec{K},k_{z};t)b^{\dagger}_{\alpha}(\vec{K},k_{z};t+\delta t)  e^{\frac{-i}{\hbar}\int_{-\infty}^{\infty}\,dt'
H^{(ext.)}(t')}\Big)|\Omega \otimes O\rangle_{c}\nonumber\\&&
\end{eqnarray}
$T$ stands for time order operator, $|O\rangle$ represents the ground state of the Hamiltonian, $H^ {(surf.)}+H^{(vac.)} +H^{(surf.-vac.)}$,    $|O\rangle_{c}$ means  $"connected"$ diagrams.

\noindent Using Wick's theorem \cite{Doniach} with respect the ground state  $|O\rangle$ we compute  the Green's function $G_{\beta,\alpha}(\vec{K},k_{z};\omega)$  to second order in the photon field.
\begin{eqnarray}
&&G_{\beta,\alpha}(\vec{K},k_{z};\omega)=\frac{i e^2v^2}{8\tilde{\epsilon}\hbar\Omega}
 \sum_{i,j=1,2}\mathbf{M_{s}(i,j|\theta,\phi)}W_{i}(\vec{K})W_{j}(\vec{K})\nonumber\\&&
g^{(\beta,c)}(\vec{K},k_{z}|\vec{K};\omega)\Big[\int\frac{d\omega_{1}}{2\pi}g^{(c,c)}(\vec{K}|\vec{K};\omega-\omega_{1})D_{s}(\omega_{1})
\Big] g^{(c,\alpha)}(\vec{K}|\vec{K},k_{z};\omega)\nonumber\\&&
\end{eqnarray}
 $\mathbf{M_{s}(i,j|\theta,\phi)}$   is the photon  matrix  polarization , $\mathbf{M_{s}(i,j|\theta,\phi)}=(\vec{e}_{s}(\theta,\phi)\cdot \vec{i})(\vec{e}_{s}(\theta,\phi)\cdot \vec{j}) $.

The Green 's function $g^{(c,\alpha)}(\vec{K}|\vec{K},k_{z};\omega)$   describes the creation of a vacuum electron with   spin $\alpha$ and the destruction of a surface electron,   $g^{(\alpha,c)}(\vec{K},k_{z}|\vec{K};\omega)$   describes inverse process  and  $g^{(c,c)}(\vec{K}|\vec{K};\omega)$  represents the Green's function for the surface electrons.  $D_{s}(\omega)$ is the photon Green's function  defined with respect the coherent  state $|\Omega\rangle$. 
\begin{eqnarray}
&&g^{(c,\alpha)}(\vec{K}|\vec{K},k_{z};t)=-i\langle O|T\Big(C (\vec{K};t)b^{\dagger}_{\alpha}(\vec{K},k_{z};0) \Big)  | O \rangle,\nonumber\\&&
g^{(\alpha,c)}(\vec{K},k_{z}|\vec{K};t)=-i\langle O|T\Big(b_{\alpha} (\vec{K},k_{z};t)C^{\dagger}(\vec{K};0) \Big)| O\rangle \nonumber\\&&
g^{(c,c)}(\vec{K}|\vec{K};t)=-i\langle O|T\Big(C (\vec{K};t)C^{\dagger}(\vec{K};0) \Big)|O\rangle\nonumber\\&&
D_{s}(t)=-i\langle \Omega|T\Big( (a_{s}(\vec{p})e^{-i\Omega t}+
a^{\dagger}_{s}(\vec{p)}e^{i\Omega t})(a_{s}(\vec{p})+
a^{\dagger}_{s}(\vec{p)})\Big)| \Omega   \rangle\nonumber\\&&
\end{eqnarray}
The Green's function $ g^{(c,\alpha)}(\vec{K}|\vec{K},k_{z};t)$  $g^{(\alpha,c)}(\vec{K},k_{z}|\vec{K};t)$ and $ g^{(c,c)}(\vec{K}|\vec{K};t)$  are obtained from the Heisenberg equations of motion,
\begin{eqnarray}
&&i\hbar\frac{ d b_{\alpha}(\vec{K},k_{z};t)}{dt}=\Big[b_{\alpha}(\vec{K},k_{z};t), H^{(vac.)} +H^{(surf.-vac.)}\Big], \nonumber\\&&  i\hbar\frac{ d C(\vec{K},z=0;t)}{dt}=\Big[C(\vec{K},z=0;t), H^ {(surf.)} +H^{(surf.-vac.)}\Big]
\nonumber\\&&
\end{eqnarray}  
We obtain:  
\begin{eqnarray} 
&&g^{(c,\alpha)}(\vec{K}|\vec{K},k_{z};\omega)=\nonumber\\&& \frac{t^{*}(k_{z})U^{*}_{\alpha}(\vec{K})}{\Big(\omega-\hat{\epsilon}+i\delta sgn[\hat{\epsilon}]\Big)\Big(\omega-\hat{E}+i\delta sgn[\hat{E}]\Big))-t^2(k_{z})}\equiv t^{*}(k_{z})U^{*}_{\alpha}(\vec{K}) \Gamma[\vec{K}|\vec{K},k_{z};\omega]\nonumber\\&&
g^{(\alpha,c)}(\vec{K},k_{z}|\vec{K};\omega)=\nonumber\\&&
\frac{t(k_{z})U_{\alpha}(\vec{K})}{\Big(\omega-\hat{\epsilon}+i\delta sgn[\hat{\epsilon}]\Big)\Big(\omega-\hat{E}+i\delta sgn[\hat{E}]\Big)-t^2(k_{z})}
=t(k_{z})U_{\alpha}(\vec{K}) \Gamma[\vec{K},k_{z}|\vec{K};\omega]\nonumber\\&&
\Gamma[\vec{K},k_{z}|\vec{K};\omega]\equiv \Gamma[\vec{K}|\vec{K},k_{z};\omega]=\frac{\Theta[\hat{\epsilon}]}{(\omega-\hat{E}_{+}+i\delta)(\omega-\hat{E}_{-}-i\delta)}+\frac{(1-\Theta[\hat{\epsilon}])}{(\omega-\hat{E}_{+}-i\delta)(\omega-\hat{E}_{-}-i\delta)}\nonumber\\&&
\end{eqnarray}
where $ \hat{E}_{\pm}=\frac{\hat{E}+\hat{\epsilon}}{2}\pm\frac{1}{2}\sqrt
{(\hat{E}-\hat{\epsilon})^2+4t^2(k_{z})}$
 are the two  roots of the algebraic equation,
 $(\omega-\hat{\epsilon})(\omega-\hat{E})-t^2(k_{z})=0$.
 $U_{\alpha}(\vec{K})$ is the  the spin  component $\alpha=\uparrow,\downarrow$ of the surface spinor  $u(\vec{K})$. $\Theta[\hat{\epsilon}]$ is the step function (one  for $\hat{\epsilon}>0$ and zero other wise) which at finite temperatures is replaced by the Fermi-Dirac occupation function.

The Green's function for the surface electrons is given by:
\begin{eqnarray}
&&g^{(c,c)}(\vec{K}|\vec{K};\omega)=
\Big[\omega-\hat{\epsilon}+i\delta sgn[\hat{\epsilon}]- \sum_{k_{z}}\frac{ t^2(k_{z})}{\omega-\hat{E}+i\delta sgn[\hat{E}]}\Big]^{-1},\nonumber\\&&
\end{eqnarray}
where $\hat{E}=\frac{\epsilon^2_{0}}{2 mv^2}+  \frac{\hbar^2 }{2m} k^2_{z}+V_{0} -\mu$.
We replace the sum by a momentum integration and use a contour integral.  The tunneling  matrix element given  $t(k_{z})$ given in eq.$(1)$ is replaced by  constant tunneling matrix element , $t(k_{z})\approx \tau\mu$, where   $\tau$ is    dimensionless ($t(k_{z})$ has dimension of energy).
The replacement of the $\sum_{k_{z}}$ in Eq.$(9)$ by the  integral $\int\,dk_{z}$ introduces the quantization length $L$. The restriction of the $z$ integration in Eq.$(9)$  by $| z| <d$ will introduce  in Eq.$(10)$ the dimensionless parameter $\frac{L}{d}$ which will be approximated by $K_{F} L$ where $K_{F}$ is the surface Fermi momentum.
\begin{eqnarray}
&&g^{(c,c)}(\vec{K}|\vec{K};\omega)=\Big[\omega-\hat{\epsilon}+i\delta sgn[\hat{\epsilon}]+i\tau ^{2} \mu^{\frac{3}{2} }( K_{F} L) \frac{\Theta[\omega-\omega_{max}[\hat{\epsilon},V_{0},\Omega]}{\sqrt{\omega-\omega_{max}[\hat{\epsilon},V_{0},\Omega]}}\Big]^{-1}\nonumber\\&&
\omega_{max}[\hat{\epsilon},V_{0},\Omega]=
(V_{0}-\mu)-\Omega+\frac{(\hat{\epsilon}+\mu)^2-\Delta^2}{2 \mu}\nonumber\\&&
\end{eqnarray}
We substitute into Eq.$(5)$ the Green's functions  $g^{(\alpha,c)}(\vec{K},k_{z}|\vec{K};\omega)$, $g^{(c,\alpha)}(\vec{K}|\vec{K},k_{z};\omega)$,   $g^{(c,c)}(\vec{K}|\vec{K};\omega)$ and perform the contour integral  with respect the photon Green's function, $\int\,d\omega_{1} g^{(c,c)}(\vec{K}|\vec{K};\omega-\omega_{1})D_{s}(\omega_{1})$,
\begin{equation}
D_{s}(\omega)=\Big[\Big(\frac{1+N_{s}}{\omega-\Omega +i\eta}+\frac{N_{s}}{\omega+\Omega +i\eta}\Big)-\Big(\frac{N_{s}}{\omega-\Omega +i\eta}+\frac{1+N_{s}}{\omega+\Omega +i\eta}\Big)\Big]
\label{photon}
\end{equation}
(In obtaining $D_{s}(\omega)$ we have used  the coherent  states properties of the photon field    with the occupation number $N_{s}$, $a^{\dagger}_{s} a_{s} |\Omega>=N_{s}|\Omega>$.)
 The  momentum integration in   Eq.$(1)$, $\int \frac{d^2 K}{(2\pi)^2}\int\,\frac{dk_{z}}{2\pi}G_{\beta,\alpha}(\vec{K},k_{z};\omega)$ is replaced in terms of the polar  angles $\chi=\phi_{d}$, $\theta_{d}$ and  the surface  energy $\epsilon$  by   $\int\, \frac{d \chi}{2\pi}\int\,\frac{dk_{z}}{2\pi}\Big[\frac{1}{(\hbar v)^2}\int\,\frac{d\hat {\epsilon}}{2\pi}\frac{(\hat{\epsilon}+\mu)}{1+\frac{\Delta}{(\hat{\epsilon}_{0}+\mu)} \frac{d\Delta}{d\hat{\epsilon}_{0}}}  G_{\beta,\alpha}(\hat{\epsilon}+\mu,k_{z};\omega)\Big]$. 
We find  from Eq. $(5)$  that the spin density measured by the detector,

$\langle \mathbf{n}^{y}(\theta_{d},\phi_{d}|\theta,\phi,s;\hat{\epsilon}+\mu) \rangle= \mathbf{P}^{y}_{s}(\theta,\phi, \theta_{d},\phi_{d},\hat{\epsilon},\Delta,\epsilon_{c})d\mathbf{I}(\theta_{d},\phi_{d},\hat{\epsilon}+\mu)$

\noindent where  $\mathbf{P}^{y}_{s}(\theta,\phi, \theta_{d},\phi_{d},\hat{\epsilon},\Delta,\epsilon_{c})$ is the spin density polarization controlled by the photon  field, $d\mathbf{I}(\theta_{d},\phi_{d},\hat{\epsilon}+\mu)$ is the number of photoelectrons per solid angle for  the surface energy 
$\epsilon$.
The  density polarization for the Zeeman gap  $ \mathbf{P}^{y}_{s}(\theta,\phi, \theta_{d},\phi_{d},\hat{\epsilon},\Delta)$ can be written as a product of  the spin density in for  the gap $\Delta=0$  (the helical state) or the warping $\epsilon_{c}$, $ \mathbf{P}^{y}_{s}(\theta,\phi,\theta_{d},\phi_{d},\hat{\epsilon}+\mu,\Delta)\equiv\mathbf{\hat{P}}^{y}_{s}(\theta,\phi,\theta_{d},\phi_{d},\Delta=0)$ and the gap function, $P_{pol.-supr.}[\hat{\epsilon}+\mu,\Delta]$. 
\begin{eqnarray}
&&\langle \mathbf{n}^{y}(\theta_{d},\phi_{d}|\theta,\phi,s;\hat{\epsilon}+\mu,\Delta) \rangle= \mathbf{P}^{y}_{s}(\theta,\phi, \theta_{d},\phi_{d},\hat{\epsilon},\Delta)d\mathbf{I}(\theta_{d},\phi_{d},\hat{\epsilon}+\mu)\nonumber\\&&
\mathbf{P}^{y}_{s}(\theta,\phi,\theta_{d},\phi_{d},\hat{\epsilon}+\mu,\Delta)\equiv\mathbf{\hat{P}}^{y}_{s}(\theta,\phi,\theta_{d},\phi_{d},\Delta=0,\epsilon_{c}=0)P_{pol.-supr.}[\hat{\epsilon}+\mu,\Delta]\nonumber\\&&
P_{pol.-supr.
}[\hat{\epsilon}+\mu,\Delta]=\Big(\frac{(\hat{\epsilon}+\mu)^2}{(\hat{\epsilon}+\mu)^2+\Delta^2}\Big)^{\frac{3}{2}}
\nonumber\\&&
d\mathbf{I}(\theta_{d},\phi_{d},\epsilon,\Delta)=
\Big(\frac{e^2 \tau^4}{4\pi\hbar \tilde{\epsilon}}\Big)( K_{F}L)
(1+N_{s})\mathbf{n}^{2}_{F}[\hat{\epsilon}]\Theta \Big[\hat{E}_{-}+\Omega-\omega_{max}[\hat{\epsilon},\Delta,V_{0},\Omega]\Big](\frac{K_{F}}{4 p_{ph.}}) \frac {\mu(\hat{\epsilon}+\mu)\sqrt{\hat{\epsilon}^2-\Delta^2}}{(\hat{E}_{+}-\hat{E}_{-})^3}\nonumber\\&&\Big(\frac{\mu^{\frac{1}{2}}} {\sqrt{\hat{E}_{-}+\Omega-\omega_{max}[\hat{\epsilon},\Delta,V_{0},\Omega]}}\Big)\Big(
 \frac{\mu^2}{(\hat{E}_{-}+\Omega-\hat{\epsilon})^2+(\tau \mu^{\frac{3}{4}})^4( K_{F} L)^2\frac{\Theta[\hat{E}_{-}+\Omega-\omega_{max}[\hat{\epsilon},\delta,V_{0},\Omega)]}{\hat{E}_{-}+\Omega-\omega_{max}[\hat{\epsilon},\Delta,V_{0},\Omega]}}\Big) 
\frac{d\phi_{d}}{2\pi}    \frac{d\theta_{d}d\hat{\epsilon}}{\pi \cos^2[\theta_{d}]} 
\nonumber\\&&
\end{eqnarray}
For the  warping  case  we have,
\begin{eqnarray}
&&\langle \mathbf{n}^{y}(\theta_{d},\phi_{d}|\theta,\phi,s;\hat{\epsilon}+\mu) \rangle= \mathbf{P}^{y}_{s}(\theta,\phi, \theta_{d},\phi_{d},\hat{\epsilon},\epsilon_{c})d\mathbf{I}(\theta_{d},\phi_{d},\hat{\epsilon}+\mu,\epsilon_{c})\nonumber\\&&
\mathbf{P}^{y}_{s}(\theta,\phi,\theta_{d},\phi_{d},\hat{\epsilon}+\mu,)\equiv\mathbf{\hat{P}}^{y}_{s}(\theta,\phi,\theta_{d},\phi_{d},\epsilon_{c}=0)P_{pol.-supr.}[\hat{\epsilon}+\mu,\epsilon_{c}]\nonumber\\&&
P_{pol.-supr.
}[\hat{\epsilon}+\mu,\epsilon_{c}]=
\Big(\frac{1}{1+(\frac{\hat{\epsilon}+\mu}{\epsilon_{c}})^4\cos^{2}(3\phi_{d})}\Big)^{\frac{3}{2}}
\nonumber\\&&
d\mathbf{I}(\theta_{d},\phi_{d},\epsilon,\epsilon_{c})=
\Big(\frac{e^2 \tau^4}{4\pi\hbar \tilde{\epsilon}}\Big)( K_{F}L)
(1+N_{s})\mathbf{n}^{2}_{F}[\hat{\epsilon}]\Theta \Big[\hat{E}_{-}+\Omega-\omega_{max}[\hat{\epsilon},\Delta,V_{0},\Omega]\Big](\frac{K_{F}}{4 p_{ph.}}) \frac {\mu(\hat{\epsilon}+\mu)\sqrt{\hat{\epsilon}^2-\Delta^2}}{(\hat{E}_{+}-\hat{E}_{-})^3}\nonumber\\&&\Big(\frac{\mu^{\frac{1}{2}}} {\sqrt{\hat{E}_{-}+\Omega-\omega_{max}[\hat{\epsilon},\Delta,V_{0},\Omega]}}\Big)\Big( \frac{\mu^2}{(\hat{E}_{-}+\Omega-\hat{\epsilon})^2+(\tau \mu^{\frac{3}{4}})^4( K_{F} L)^2\frac{\Theta[\hat{E}_{-}+\Omega-\omega_{max}[\hat{\epsilon},\delta,V_{0},\Omega)]}{\hat{E}_{-}+\Omega-\omega_{max}[\hat{\epsilon},\Delta,V_{0},\Omega]}}\Big)\nonumber\\&&
\frac{1}{1+3(\frac{\hat{\epsilon}+\mu}{\epsilon_{c}})^{4}\cos^2[3\phi_{d}]}\frac{d\phi_{d}}{2\pi}    \frac{d\theta_{d}d\hat{\epsilon}}{\pi \cos^2[\theta_{d}]}
\nonumber\\&&
\end{eqnarray}
$\mathbf{n}_{F}[\hat{\epsilon}]$ is the Fermi Dirac occupation function and $\Theta \Big[x\Big]$ is the step function  which is one for $x>0$ and zero otherwise. $d\mathbf{I}$  represents the number of emitted  photoelectrons per solid angle $\theta_{d}$,$\phi_{d}$ and  surface energy $\epsilon$. This number depends on the number of the  incoming  photons $N_{s}$, the ratio  between the Fermi momentum $K_{F}$  and the photon momentum $p_{ph.}=\frac{\Omega}{c}$  the  effective charge $e\cdot\tau^2$ and chemical potential $\mu$.
We have plot  the function $d\mathbf{I}$ for the  followings   values:

$\mu= 0.1 eV$  is the chemical potential,$V_{0}=5 eV$ is the binding energy 
$\Omega=5 eV$ is the laser frequency, $\tau=0.1$ is the dimensionless  tunneling parameter,  $K_{F}L=10^{7}$ is determined by the distance from the surface sample to the detector,
 $v=5\times  10^5$ $\frac{m}{sec}$ is the Fermi surface velocity 
,$\frac{e^2}{4\pi\hbar c \tilde{\epsilon}}=\frac{1}{137}$ is the fine structure constant  and  $N_{s}=10^{12}$ is  the number of photons. The plots will be restricted to energies $\hat{\epsilon}\geq-\mu$ which represents  the surface electrons, the bulk contributions will be ignored.

We find  that  $10^{12}$ photons are needed  for two  photoelectrons to be emitted.
Figure $1$, $2$ , $3$ and   $4$  shows the number of photoelectrons emitted  for  $\Delta=0$,  $\Delta=0.05 eV$,  warping energy   $\epsilon_{c}=0.05 eV$,$\phi_{d}=\frac{\pi}{4}$ and $\phi_{d}=\frac{\pi}{6}$.  In all the four cases we  have used the same values. The   warping the angle $\phi_{d}$ controls the number of   the  photoelectrons . 
\begin{figure}
\begin{center}
\includegraphics[width=2.5 in ]{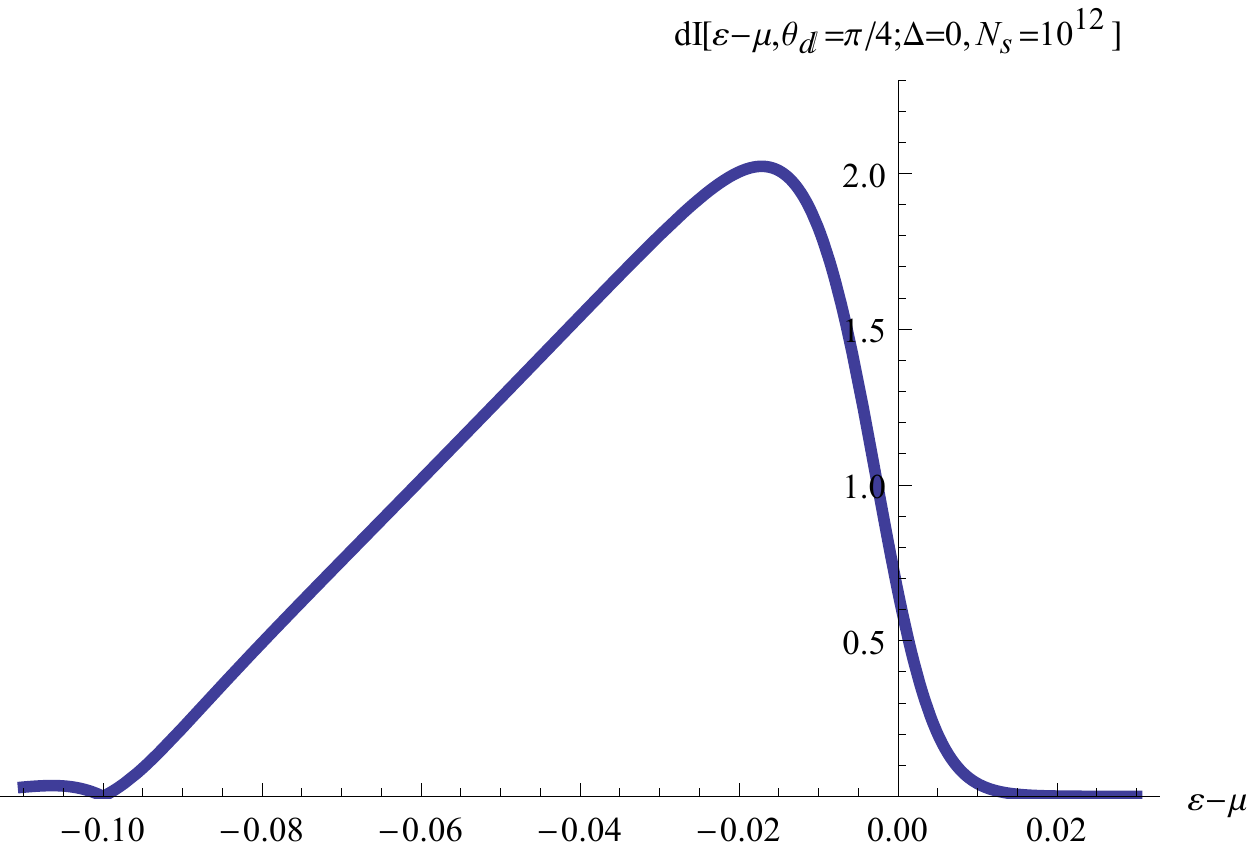}
\end{center}
\caption{The number of photoelectrons $ d\mathbf{I}(\theta_{d},\phi_{d},\epsilon)$ for $\Delta=0$}
\end{figure} 
\begin{figure}
\begin{center}
\includegraphics[width=3.0 in ]{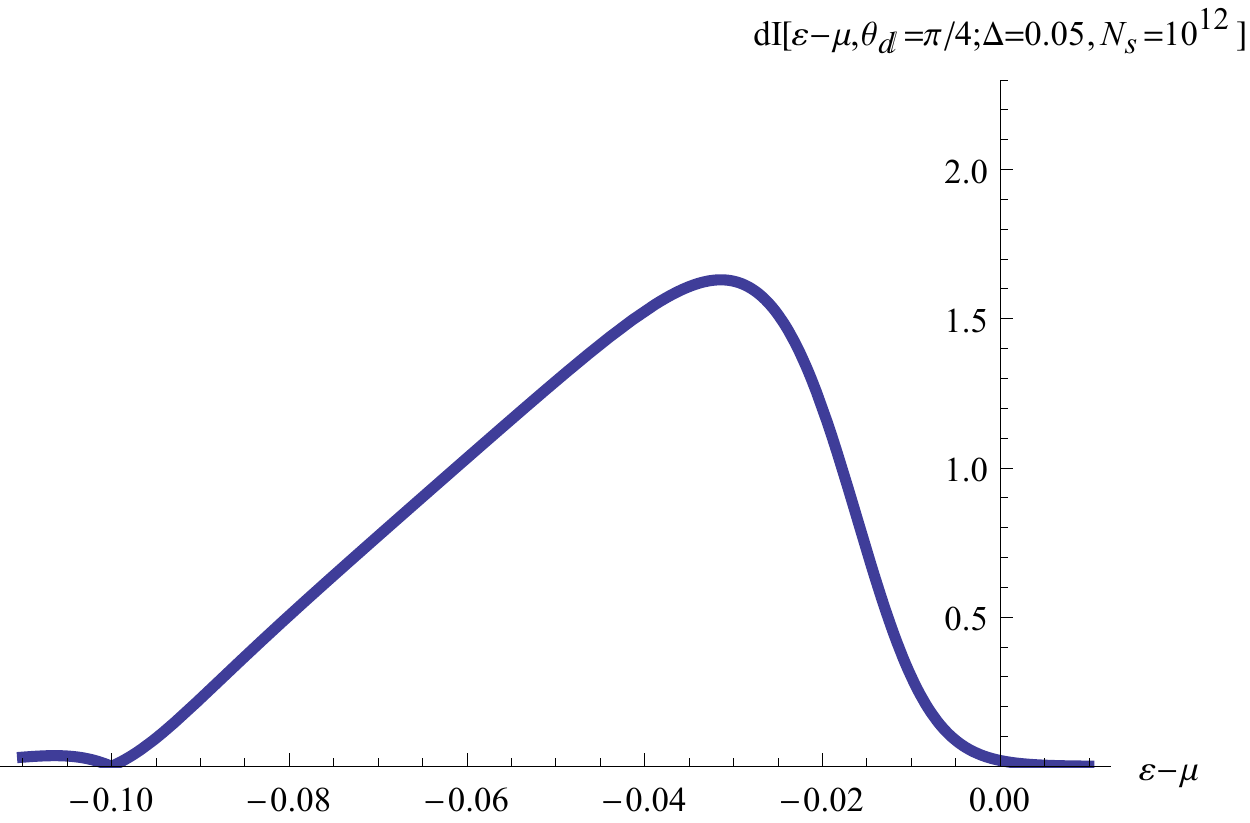}
\end{center}
\caption{The number of photoelectrons $ d\mathbf{I}(\theta_{d},\phi_{d},\epsilon)$ for  $\Delta=0.05 eV$} 
\end{figure}
\begin{figure}
\begin{center}
\includegraphics[width=3.0 in ]{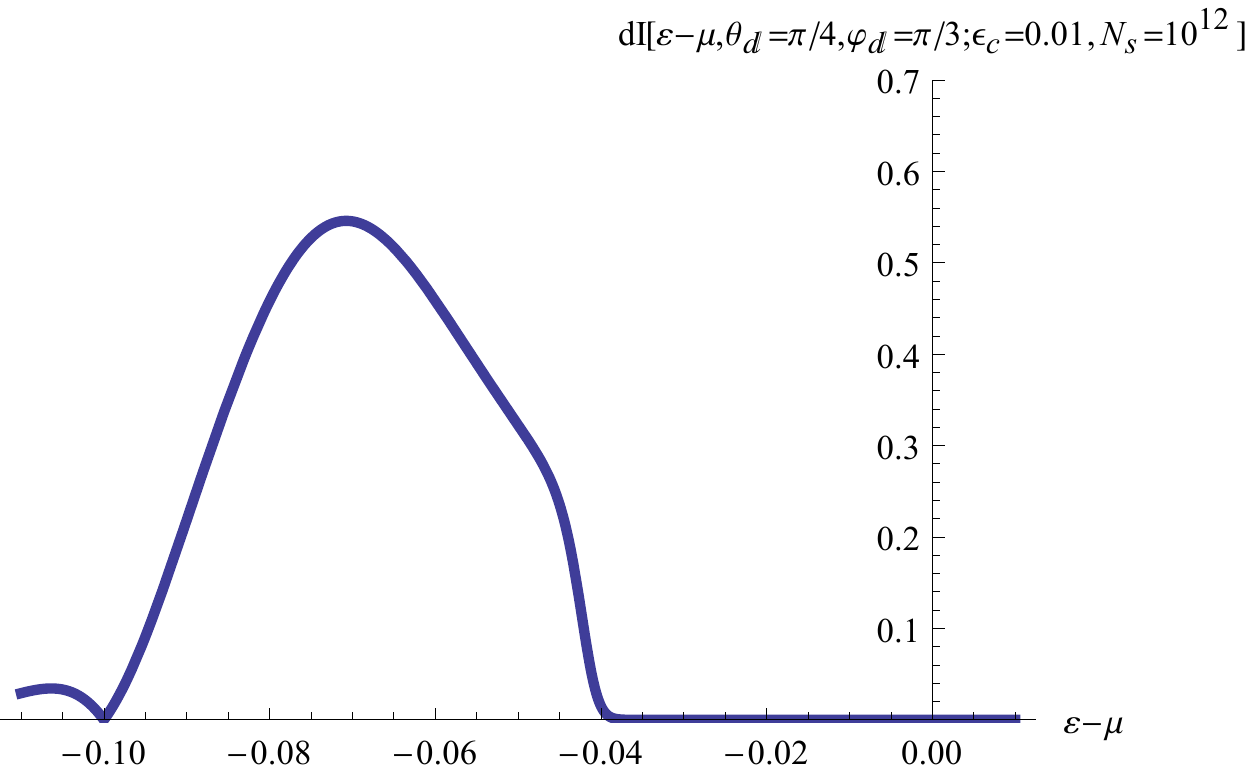}
\end{center}
\caption{The number of photoelectrons $ d\mathbf{I}(\theta_{d},\phi_{d},\epsilon)$ for  $\epsilon_{c}=0.05 eV$, $\phi_{d}=\frac{\pi}{3}$} 
\end{figure}
\begin{figure}
\begin{center}
\includegraphics[width=3.0 in ]{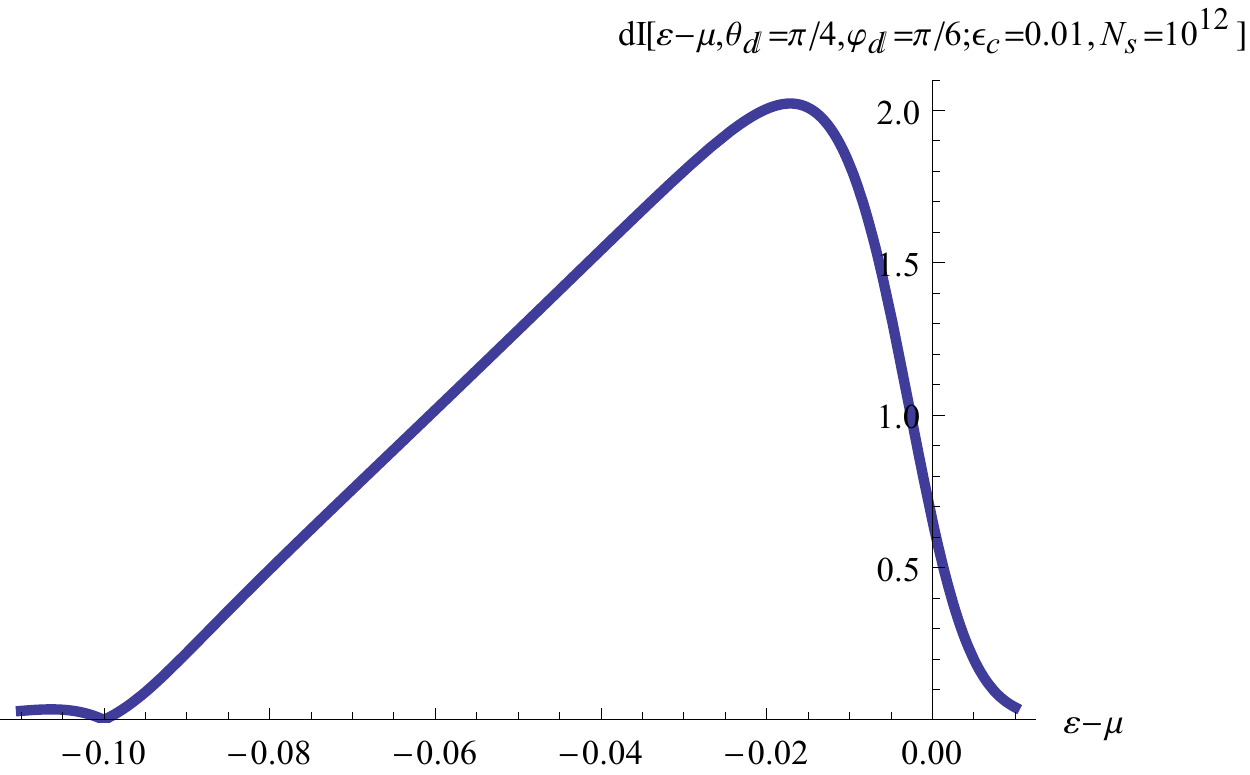}
\end{center}
\caption{The number of photoelectrons $ d\mathbf{I}(\theta_{d},\phi_{d},\epsilon)$ for  $\epsilon_{c}=0.05 eV$, $\phi_{d}=\frac{\pi}{6}$} 
\end{figure}
Each time the angle $\phi_{d}= (2n+1)\frac{\pi}{6}$, $n=0,1,2,..$ the intensity is maximum and is minimum when  
the angles are  $\phi_{d}= (2n+1)\frac{\pi}{3}$.

\vspace{0.2 in}

\textbf{IV- The spin density in the $y$ direction   measured by the detector at the polar angle $\phi_{d}$,$\theta_{d}$ as a function of the incoming  photons $\vec{e}_{s=1,2}(\theta,\phi)$}

\vspace{0.2 in}

The spin polarization of   the  $TI$ surface is given in terms of the spinor states $U_{\alpha}(\vec{K})$ with $\alpha=\uparrow,\downarrow$.
The momentum parallel to the surface is conserved, the chiral angle satisfies $\cos[\phi_{d}]=\cos[\chi]$.
The spin polarization
 $ \langle\sigma^{y}\rangle\equiv S^{y}=\cos[\phi_{d}](\sin[\beta[\hat{\epsilon},\Delta,\epsilon_{c}])^\frac{3}{2}\Theta[\hat{\epsilon}]$ 
  is recorded by the detector as   $\mathbf{\hat{P}}^{y}_{s}(\theta,\phi,\phi_{d},\hat{\epsilon})$. $\sin[\beta[\hat{\epsilon},\Delta,\epsilon_{c}]$ represent the  effect of the Zeeman gap $\Delta$ or ,$\epsilon_{c}$   for warping. The  function $(\sin[\beta[\hat{\epsilon},\Delta,\epsilon_{c}])^\frac{3}{2} $ is given for  the gap $\Delta$ and  warpping  $ \epsilon_{c}$,
\begin{eqnarray}
&&P_{pol.-supr.
}[\hat{\epsilon}+\mu,\Delta]\equiv( \sin[\beta[\hat{\epsilon},\Delta)])^{\frac{3}{2}}=\Big(\frac{(\hat{\epsilon}+\mu)^2}{(\hat{\epsilon}+\mu)^2+\Delta^2}\Big)^{\frac{3}{2}}\nonumber\\&&
P_{pol.-supr.
}[\hat{\epsilon}+\mu,\epsilon_{c}]\equiv(\sin[\beta[\hat{\epsilon},\epsilon_{c})])^\frac{3}{2}=\Big(\frac{1}{1+(\frac{\hat{\epsilon}+\mu}{\epsilon_{c}})^4\cos^{2}(3\phi_{d})}\Big)^{\frac{3}{2}}\nonumber\\&&
\end{eqnarray}
 Using the definition given in Eqs.$(4-6)$ we obtain the relation between $ S^{y}$ and    $\mathbf{\hat{P}}^{y}_{s}(\theta,\phi,\phi_{d}\hat{\epsilon},\Delta=\epsilon_{c}=0)$:
\begin{eqnarray}
&& \mathbf{\hat{P}}^{y}_{s}(\theta,\phi,\phi_{d},\hat{\epsilon},\Delta=\epsilon_{c}=0)=S^{y}\sum_{i,j=1,2}\mathbf{M_{s}(i,j|\theta,\phi)}W_{i}(\phi_{d},\hat{\epsilon} )W_{j}(\phi_{d},\hat{\epsilon} ) W_{i}(\phi_{d},\hat{\epsilon} )\nonumber\\&&
\end{eqnarray}
The spin density for  the linear for the $\mathbf{p}$  polarization polarization  corresponds to $\vec{e}_{s=1}(\phi=0,\theta)$,and  the $\mathbf{s}$ polarization  is corresponds to  $\vec{e}_{s=2}(\phi=0)$.
We find: 
\begin{eqnarray}
&&\mathbf{\hat{P}}^{y}_{s=1}(\theta,\phi=0,\phi_{d},\Delta=\epsilon_{c}=0)=S^{y}  \cos^2[\theta] \cos^2[\phi_{d}]\nonumber\\&& 
\mathbf{\hat{P}}^{y}_{s=2}(\phi=0,\phi_{d},\Delta=\epsilon_{c}=0)=S^{y}  \sin^2[\phi_{d}]\nonumber\\&&
\end{eqnarray}
 The detection of the spin polarization  is affected  by the photon polarization. We have the relation: 
\begin{equation}
\mathbf{\hat{P}}^{y}_{s=2}(\frac{\pi}{2}-\phi_{d},)\cos^2[\theta]=\mathbf{\hat{P}}^{y}_{s=1}(\theta,\phi_{d})
\label{eqt}
\end{equation}
In figure $5$  shows the polarizations for $\Delta=\epsilon_{c}=0$. $\mathbf{\hat{P}}^{y}_{s=1}(\theta_{d}=\frac{\pi}{4},\phi_{d})$ represents  the $\mathbf{p}$ polarization  and $\mathbf{\hat{P}}^{y}_{s=2}(\theta_{d}=\frac{\pi}{4},\phi_{d})$ is the result for the  $\mathbf{p}$ polarization. 
\begin{figure}
\begin{center} 
\includegraphics[width=3.0 in ]{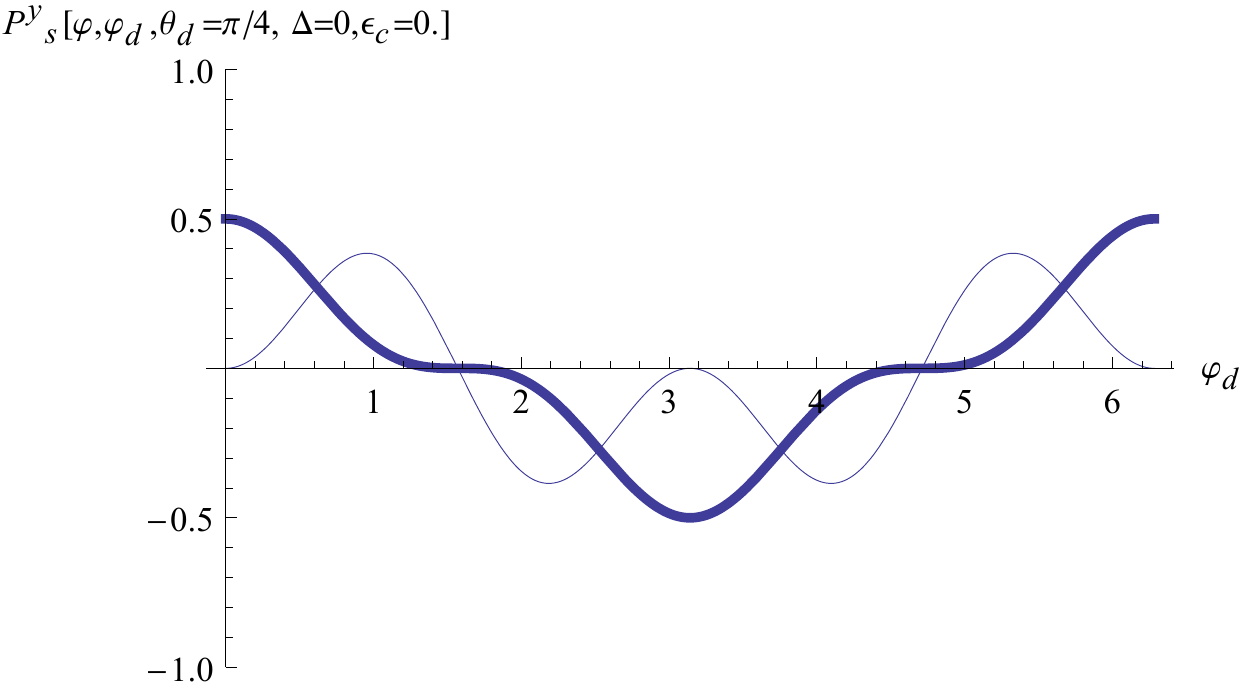}
\end{center}
\caption{$\mathbf{\hat{P}}^{y}_{s}(\theta_{d}=\frac{\pi}{4},\phi_{d})$ for $\Delta=\epsilon_{c}=0$. The thick line represents the $\mathbf{p}$  polarization   $\mathbf{P}^{y}_{s=1}$ and   the thin line represents the  $\mathbf{s}$ polarization $\mathbf{P}^{y}_{s=2}$ } 
\end{figure} 

For a finite gap $\Delta$  the polarization is suppressed by  the  energy factor $(\sin[\beta[\hat{\epsilon},\Delta)])^\frac{3}{2}$.

For warping  figure $6$  shows the warping caused by  the angle  $\phi_{d}$, $(\sin[\beta[\hat{\epsilon},\epsilon_{c})])^\frac{3}{2}$. 
\begin{figure}
\begin{center}
\includegraphics[width=3.0 in ]{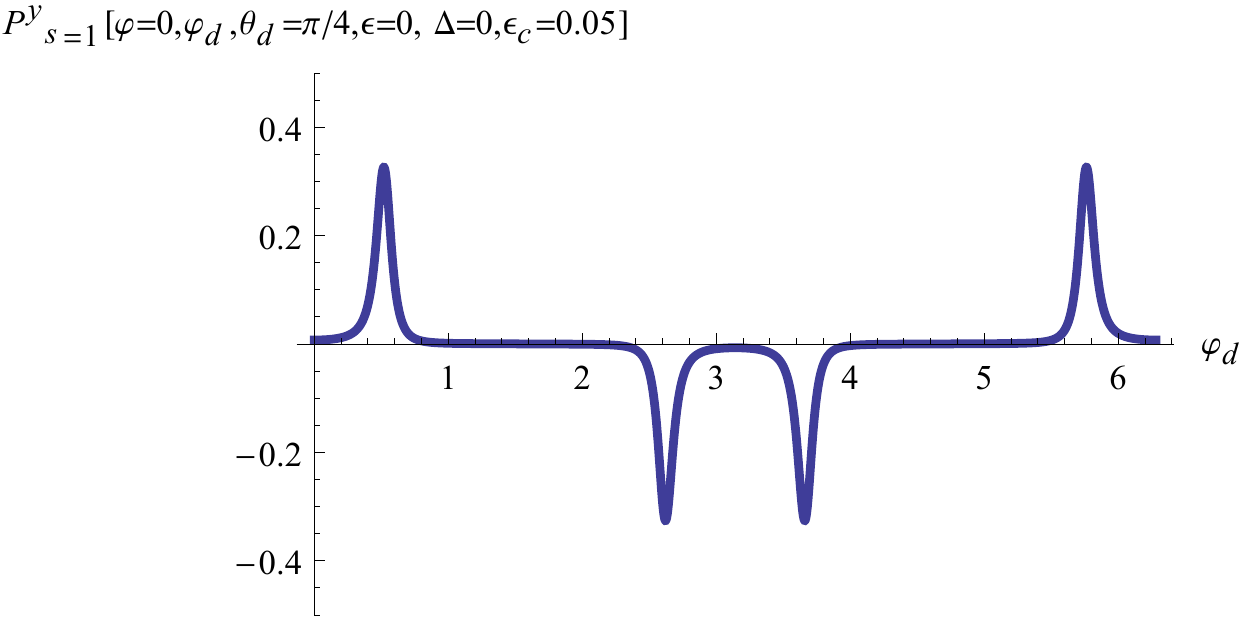}
\end{center}
\caption{$\mathbf{\hat{P}}^{y}_{s=1}(\theta_{d}=\frac{\pi}{4},\phi_{d})$ for $\epsilon_{c}=0.05$ at the  fixed energy. } 
\end{figure}

\vspace{0.2 in}

\textbf{V-Conclusions}

\vspace{0.2 in}

To conclude, a model  for   computing  the photoelectrons intensity  and polarization from the surface of a topological insulator  based on Green's functions  has been introduced.  We  show that the polarization of  photoelectrons depends on the laser light polarization in  qualitative agreement with experimental results \cite{Nature}.  The photoelectrons polarization  is 
modified by  the incoming  photon polarization.
This results hold also in  the presence of a Zeeman gap or warping. For the Zeeman gap the polarization and intensity  is suppressed  closed  to the Fermi  energy. For warping the  polarization and the  intensity oscillates with the warping angle $\phi_{d}$ allowing to identify  the spin texture.

The calculation is based on the tunneling amplitude of the surface 
electrons into the  vacuum .This amplitude can be estimated from the inverted and vacuum $TI$   gap.  We compute the number of the emitted  photoelectrons.  This number depends tunneling amplitude, chemical potential,incoming number of photons  and weakly dependent on the location of the detector.
Our calculation indicate that for $\tau=0.1$, $\mu=0.1 eV$,  $K_{F}L=10^{7}$ and $N_{s}=10^{12}$ we obtain that the maximum number of photoelectrons  is approximately $2$.

The calculation ignores the bulk electrons, therefore  the intensity  computed can not be compared with the experimental results for energy $\epsilon-\mu<-\mu$ (the bottom of the surface states are at $\epsilon_{0}=\hbar v|\vec{K}|=0$ , in figures $1-4$ this corresponds to  $\epsilon-\mu=-0.1eV$ )

\end{document}